\documentclass{article}
\usepackage[utf8]{inputenc}
\usepackage{graphicx}

\title{Impact of Dietary Habits and Opinionated Lifestyle during COVID-19 Pandemic : A Case Study on Engineering Students}
\author{Arpitha A. Deshpande ( arpitaadeshpande07@gmail.com ) \and
Aadrika A. ( aadrika1916@gmail.com ) \and
Rajeshwari K. ( rajeshwarik.ise@bmsce.ac.in ) \and
Preetha S. ( preetha.ise@bmsce.ac.in ) }

\date{}
\begin{document}

\maketitle
\begin{abstract}
    COVID-19 pandemic has introduced a new lifestyle due to lockdown. The impact was on food habits, working hours, and sleeping patterns.  The goal of this study is to detect lifestyle changes caused by confinement during the COVID-19 pandemic, such as dietary habits, physical activities, and to explore changes in the body weight. A structured questionnaire was used in the study to collect anthropometric data; daily consumption of particular foods, water intake, food frequency, and number of meals/day. The data is presented in a graph illustration to show health , lifestyle trends and the friendship among engineering students. 


\end{abstract}
\section{Introduction}
The COVID-19 epidemic has presented a new set of obstacles for the individuals to maintain a healthy diet. Many countries saw the lockdown to curb the pandemic. Countries imposed new rules to curb the spread of the disease, people were not allowed to step out of their homes, until necessary. The restriction of one's activities has a direct impact on their lifestyle, including food choices, eating habits, and physical activity patterns. As individuals are locked up, boredom bothers them more. Boredom has been linked to a higher calorie intake as well as increased fat, carbohydrate, and protein consumption. People are compelled to overeat as a result of their stress, primarily in search of sweet "comfort foods." Also, a mix of working from home, online schooling, and social media usage resulted in an increase in screen time. Such factors have led to weight fluctuations in a very small time period. To safeguard metabolic adaptations, minimise systemic inflammation, and enhance nutritional practises that might offset the consequences of confinement, it is extremely important to maintain healthy eating habits and fitness during the COVID-19 pandemic. A survey was conducted in B.M.S. College of Engineering for 6th semester students of the Information Science and Engineering branch. Responses were collected to know whom a student would name as their "Fitness-Buddy" and the similarities between individuals and their friends was examined. As a result,  the study identified changes in lifestyle caused by confinement during the COVID-19 pandemic.

\section{Literature Survey}
Arzu zlem et al. \cite{1} explored how people's dietary habits changed during the COVID-19 pandemic lockdown. A systematic review was carried out by searching the PubMed database and Google Scholar for publications published between March and July 2020; a total of nine articles were evaluated. The lockdown had both good and bad consequences on eating patterns, such as a switch to home meals and a decrease in fast food intake, but eating frequency has risen owing to stress. Consuming immune-boosting foods, meal times, and quantities with positive thinking  can alleviate the bad health impacts of pandemic. Changes in eating habits induced by increased stress ows to increased food consumption and a snacking inclination. Joseph Godefroy et al. \cite{2} study illustrates how individual fitness influencers and organisations used Instagram during the pandemic to contribute knowledge to feed debates about the effects of COVID-19 on sport and active living. The study observed that fitness influencers encouraged physical activity at home by examining the content published during the COVID-19 outbreak. Online physical training was more accessible than ever during this time period. The diversity of the content offered and delivered to Instagram users played an important role in providing opportunities for people to practice sport at home during lockdown. During the COVID-19 pandemic, Daniela Reyes-Olavarra et al. \cite{3} sought to discover lifestyle changes caused by confinement. The poll included 700 Chilean national territory inhabitants ranging from 18 to 62 (women, n = 528, and men, n = 172). A study was conducted in May and June 2020 to evaluate eating habits, physical activity, body weight, and sociodemographic factors. Body weight gain was connected to three times per week intake of fried foods, lack of water, and six-hour daily inactive period. Fish eating, active breaks, and four hours of physical exercise each week were all found to have an inverse relationship with body weight increase. Alcohol intake on a daily basis was linked to a reduction in physical activity. Food choices, physical activity, and active breaks influence in avoiding weight gain during COVID-19 confinement. The study presented many and pertinent factors that are to be addressed during COVID-19. Creation of active breaks and physical activity helps in balanced body weight.

Laura Di Renzo et al. \cite{4} investigated the early effects of the COVID-19 pandemic on food habits and lifestyle changes in a 12-year-old Italian community. A standardised questionnaire package was used in the study to collect demographic data, anthropometric data, dietary habits data, and lifestyle habits data. The study included 3533 people ranging in age from 12 to 86 years (76.1 percent females). Weight gain was seen by 48.6 percent of the population; 3.3 percent of smokers opted to quit smoking; 38.3 percent of respondents reported a small increase in physical activity, particularly bodyweight training. When compared to the younger and elderly populations, the population aged 18–30 years had a higher adherence to the Mediterranean diet (p0.001; p0.001, respectively); 15\% of respondents turned organic, purchasing fruits and vegetables, especially in the North and Center of Italy, where BMI values were lower. 

In \cite{5}, Michelle et al.,  conducted a study on dietary pattern and behaviour of brazilian children and adolescents during COVID-19  pandemic. Social isolation influenced eating habits by enabling non-isolated families to consume healthy food less frequently. Dietary patterns played a key role in protecting individuals from the severity of COVID-19 and other infections. Contextual factors for such behavior were due to personal reasons, loss of income, anxiety related to COVID-19, gender and household composition. The study helped in identifying populations that are particularly vulnerable to nutritional changes during the pandemic, and potential avenues that could be explored to minimize the negative effects of the pandemic on food intake in consumers.
Emily W. Flanagan et al. \cite{7} collected data on dietary habits, physical activity, and mental health in an online survey for adults in April 2020. All questions were phrased "before" and "since" the COVID-19 pandemic. In total, 7,753 people were included in the study; 32.2 percent were normal weight, 32.1 percent were overweight, and 34.0 percent were grossly overweight. Overall scores for healthy eating increased (P<0.001) during the pandemic, owing to less eating out and preferred home food (P<0.001). Duration spent in physical activity (absolute time and intensity adjusted) decreased, while sedentary leisure behavior increased (P<0.001). During the pandemic, anxiety scores increased by 8.78±0.21, and the amount of the rise was significantly greater among persons who were overweight (P<0.01). Weight gain was reported by 27.5 percent of the whole group, versus 33.4 percent of obese participants. The fear of getting the virus had a substantial impact on lifestyle patterns, as well as mental health declines. Obese people have been disproportionately affected by these negative consequences. 

Stephanie Stockwell et al. \cite{8} intended to summarise research that studied changes in physical activity and sedentary behaviour before and during the COVID-19, due to the multiple health consequences related with physical activity and sedentary behaviour. Sixty-six papers (total n=86 981) satisfied the inclusion criteria and were included in the review. Physical activity changes were documented in 64 research studies, with the majority of studies showing declines in physical activity and increases in sedentary behaviour across different groups, including children and patients with vivid medical problems. Given the numerous physical and mental benefits of increased physical activity and decreased sedentary behaviour, public health strategies should include the development and implementation of interventions that promote safe physical activity and reduce sedentary behaviour in the event of other lockdowns.

Roberta Zupo et al. \cite{9}, investigated the preliminary impacts of the lockdown lifestyle on food choices. Preliminary data indicated a significant increase in the consumption of carbs, particularly those with a high glycemic index i.e. homemade pizza, bread, cake, and pastries, as well as more frequent snacking. A high intake of fruits and vegetables, as well as protein sources, notably pulses, was observed, but there was no discernible peak in the latter. In the extremely realistic scenario of a continuous global health emergency represents a serious challenge. Nutrition is a priority as eating a well-balanced diet is the greatest way to obtain all the necessary nutrients to maintain proper immunological function while lowering the risk of obesity. Furthermore, it would be preferable to give particular training for health professionals to enable them to help individuals living with obesity in the COVID-19 period. Therefore to reduce associated dangers and unconscious prejudices, as well as to avoid stigma, a balanced diet need to be followed.

\section{Dataset Description}
The data was collected via a survey through Google forms. The poll was primarily concerned with the students' eating habits, physical activity routines, and fitness associations. In college premises, students are engaged in classroom activities throughout the day with transitioning between laboratories, classroom and campus. Virtual classes have increased the screen time along with seated time and causes less mobility. Thus the daily routine has a huge change with less time constraint
\subsection{Food Habits}
A daily consumption frequency survey was utilised to collect antecedents connected to eating. In \cite{6}, a survey was conducted to know the food consumption habits and their frequency. Based on the demographic characteristics, food habits varied. This technique entailed obtaining as much information as possible regarding the frequency of :
\begin{itemize}
    \item Daily meal consumption (less than 3, exactly 3 or greater than 3)
    \item Type of meal consumed (high in proteins, fats, carbohydrate or comparatively a balanced diet) 
    \item Fruits and vegetables consumed per day (1, 2, 3 or more servings a day) \item Daily water intake (less that 3ltr, 1-3ltr or more than 3ltr)
\end{itemize}
In terms of behavioural changes compared to prior COVID-19 confinement, questions were asked regarding food habits and weight variations.
\subsection{Physical Activity Patterns}
The number of workout hours per week was investigated (none, less than 3hrs, 3-6 hrs, or more than 6hrs). They were also questioned about the friends they wish to work out with. Sleeping patterns (less than 5 hours, 5-7 hours, 7-9 hours, or more than 9 hours) were also taken into account since they contribute a significant influence in a person's ability to be active or stressed.  Each student was required to mention at least one friend’s name as  Fitness-Buddy whom they wish to workout with. Homophily is defined as people’s “tendency to associate with others whom they perceive as being similar to themselves in some way” \cite{10}. The existence of homophily at a dyadic level was observed, students chose peers with similar characteristics as  their Fitness-Buddy.

\section{Inferences}
The purpose of this study is to determine the positive and negative effects of the COVID-19 pandemic on the dietary habits of adolescents. It discusses the impact of the epidemic on eating patterns and raises awareness of the issue. Furthermore, the findings may aid in the design and targeting of programmes to provide adequate guidance for optimizing health during and after the COVID-19 outbreak by eating a balanced diet, staying hydrated, being physically active, getting enough sleep, and managing stress. 

\begin{figure}[h]
       \centering
       \includegraphics[width=\linewidth]{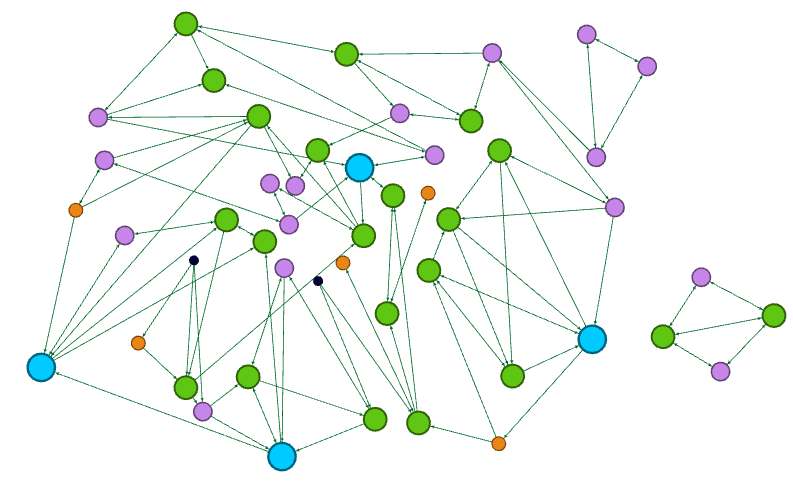}
       \caption{Fitness-Buddy Friendship Graph}
\end{figure}

The network shown in figure 1 was developed in Gephi using the “Yifan Hu” layout. The network has 49 nodes and 118 directed edges. This network focuses on student friendship with an average degree of 2.612, which implies that on an average, one person is related to two individuals. Out of 49 students, the node with the greatest in-degree is deemed the most chosen. Fitness-Buddy graph properties are listed in table 1. Fitness-buddy friendship graphs can evolve to build new friendships, as there is a shared interest which promotes workout times and improving  health. As the friendship network expands, there is scope for learning new skills and knowledge about new sports or activities is enhanced.
\begin{table}[ht]
    \centering
    \caption{Fitness-Buddy graph properties}
    \begin{tabular}{|c|c|}
    \hline
    Property & Fitness-Buddy Friendship Graph \\
    \hline
    Graph Type & Directed \\
    \hline
    Nodes & 49 \\
    \hline
    Edges & 118 \\
    \hline
    Unique directed edges & 26 \\
    \hline
    Bi directional edges & 46 \\
    \hline
    Eccentricity & 16 \\
    \hline
    Harmonic Closeness Centrality & 1.0 \\
    \hline
    Eigenvector Centrality & 1.0 \\
    \hline
    \end{tabular}
    \label{tab:my_label}
\end{table}

Figure 1 also shows the nodes with the maximum in-degree. The maximum in-degree of the nodes StudID9, StudID23, StudID29, StudID31 is 4. The edges indicating in-degree are blue, whereas the edges defining out-degree are pink. The size and colour of the nodes in the graph are varied by their in-degrees. Blue nodes have the highest in-degree, i.e. 4. The in-degrees of green, purple, orange, and black nodes are 3, 2, 1, and 0, respectively. Based on in-depth analysis, Table 2 describes overall commonalities in eating habits and physical activity of most influential nodes. Table 3 defines the node with minimum in-degree of 0, indicating that none of the other participants chose them. As a result, no parallels could be drawn.

\begin{table}[ht]
\centering
\caption{Nodes and their Similarities}
\resizebox{\textwidth}{!}{\begin{tabular}{|c|c|c|c|c|c|c|}
  \hline
  Node & In-degree & Hub Score & Authority Score & Page Rank & Friends & Similarities  \\ 
  \hline
    StudID9 & 4 & 0.000752 & 0.033004 & 0.031826 & StudID14 &  \\
     & & & & & StudID18 & Servings of fruits \\
     & & & & & StudID34 & \\
     & & & & & StudID41 & \\
     \hline
     StudID31 & 4 & 0.003094 &0.010011 & 0.025056 & StudID29 &  Sleeping hours\\
     & & & & & StudID30 &common friend \\
     & & & & & StudID38 &meal count and type \\
     & & & & & StudID43 &obese \\
     \hline
     StudID29 & 4 & 0.007598 & 0.010219 & 0.023753 & StudID2 &  Common friend\\
     & & & & & StudID4 &sleeping hours \\
     & & & & & StudID15 &meal count\\
     & & & & & StudID39 &obese \\
     \hline
     StudID23 & 4 & 0.226948 & 0.588436 & 0.027472 & StudID16 &  \\
     & & & & & StudID27 &None \\
     & & & & & StudID33 & \\
     & & & & & StudID36 &\\
   \hline
\end{tabular}}
\end{table} 

\begin{table}[ht]
\centering
\caption{Nodes and their Similarities}
\resizebox{\textwidth}{!}{\begin{tabular}{|c|c|c|c|c|c|c|}
  \hline
  Node & In-degree & Hub Score & Authority Score & Page Rank & Friends & Similarities  \\ 
  \hline
    StudID26 & 0 & 0.000555 & 0.0 & 0.003061 & None & None\\
    \hline
    StudID13 & 0 & 0.015726 & 0.0 & 0.003061 & None & None\\
   \hline
\end{tabular}}
\end{table} 

\subsection*{}
\subsection{Statistical Visualization}
  \begin{figure}[h]
       \centering
       \includegraphics[scale=0.2]{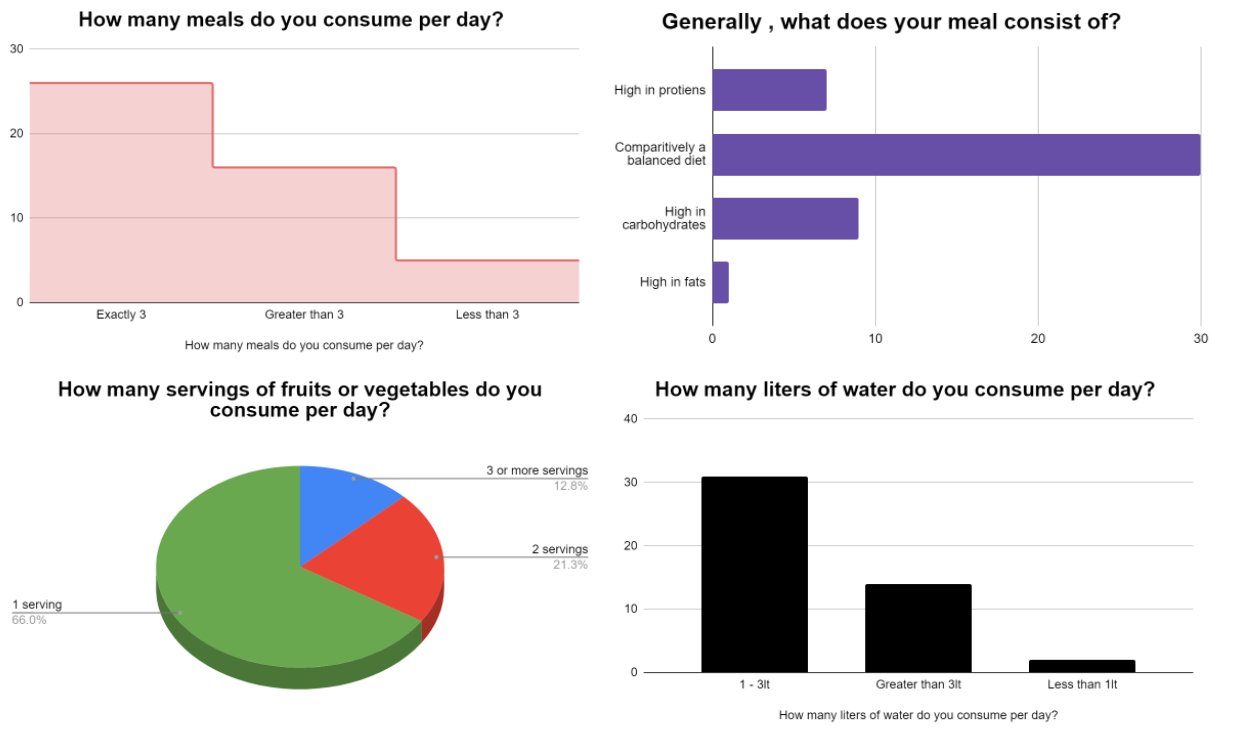}
       \caption{Dietary habits}
  \end{figure}

The dietary habits chart of the students are depicted in figure 2. The most popular patterns were the meal count of exactly three times a day,, fruits and vegetable servings of one, a relatively balanced diet as a diet pattern, and water consumption of 1-3 litres, as per the responses.

\begin{figure}[h]
        \centering
        \includegraphics[scale=0.2]{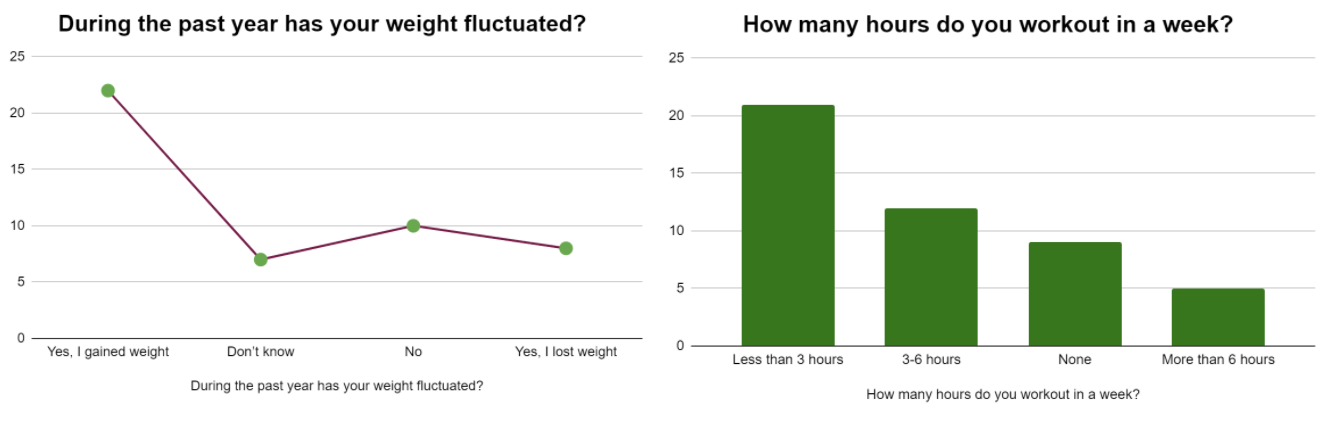}
        \caption{Workout and weight gain}
\end{figure}

Figure 3 shows that the majority of students work out for less than three hours per week, which has contributed to the weight gain throughout the pandemic in some way.

\begin{figure}[h]
        \centering
        \includegraphics[scale=0.25]{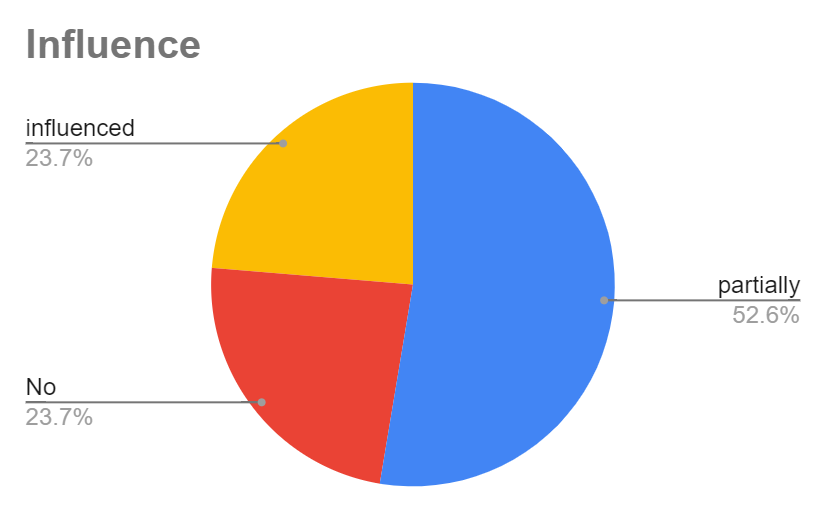}
        \caption{ Influence factor}
\end{figure}
\subsection*{}
Figure 4 depicts the extent to which students were impacted by their friends in terms of weight increase or reduction. Most students were partially impacted in terms of weight gain or loss by following their friends' suggestions in terms of eating habits, while others did not.

\begin{figure}[h]
        \centering
        \includegraphics[scale=0.3]{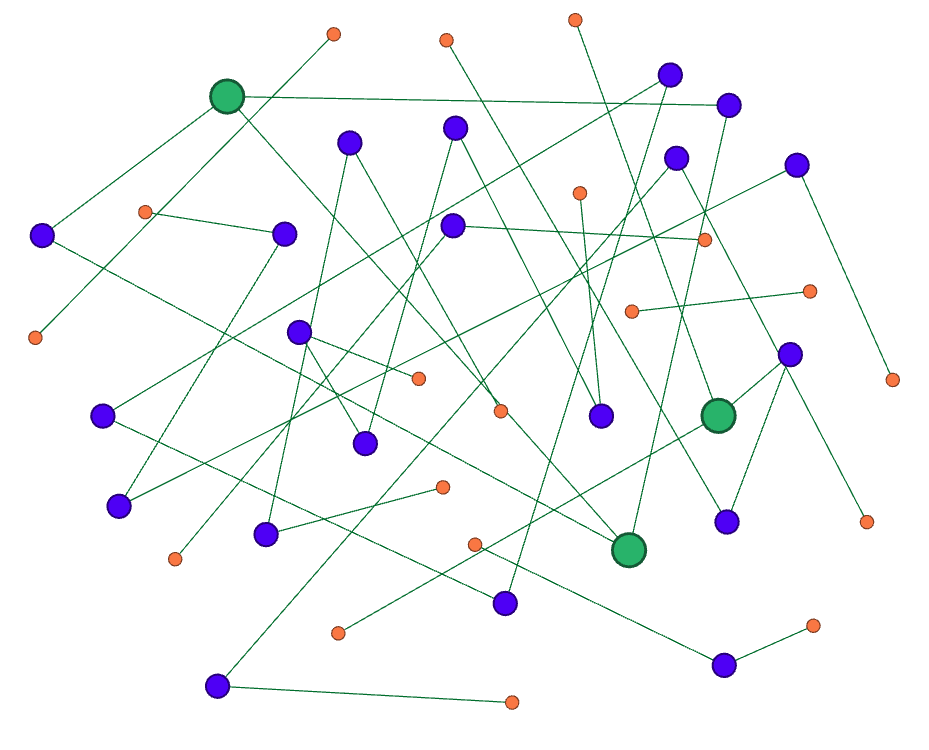}
        \caption{Fitness-Buddy Friendship Graph-Undirected}
\end{figure}

Figure 5 is created by bi-directional edges of the graph of  figure 1. The fitness-buddy suggests which friend is chosen to be a buddy in workout. If two friends choose each other preferably for working out together, their characteristics are more similar to each other. Friendships during workouts are more adventurous as one gets motivated by other individuals, leading to consistency. The size and colour of the nodes in the graph are determined by their degrees. The degrees of green, dark blue and orange nodes are 3, 2 and 1 respectively. Table 4 gives the properties of  Undirected Fitness-Buddy Graph.

\begin{table}[ht]
    \centering
    \caption{ Characteristics of undirected Fitness-Buddy Graph}
    \begin{tabular}{|c|c|}
    \hline
    Property & Fitness-Buddy Friendship Graph-Undirected \\
    \hline
    Graph Type & Undirected \\
    \hline
    Nodes & 42 \\
    \hline
    Edges & 34 \\
    \hline
    Number of triangles & 19 \\
    \hline
    Eccentricity & 5 \\
    \hline
    Harmonic Closeness Centrality & 1.0 \\
    \hline
    Eigenvector Centrality & 1.0 \\
    \hline
    \end{tabular}
    \label{tab:my_label}
\end{table}

\section{Discussions}
Table 5 depicts the comparison of directed and undirected graph of Fitness-Buddy data. An observation is made on the graph properties where network diameter and graph density decrease as the edge count decreases. The undirected graph in figure 5 illustrates more  strongly connected components than the directed graph  in figure 1.

\begin{table}[ht]
\centering
\caption{Comparison  of Fitness-Buddy graphs}
\resizebox{\textwidth}{!}{\begin{tabular}{|c|c|c|}
  \hline
  Property & Fitness-Buddy Friendship Graph & Fitness-Buddy Friendship Graph-undirected \\ 
  \hline
   Graph Type & Directed & Undirected\\
   \hline
   Nodes & 49 & 42\\
   \hline
   Edges & 118 & 34\\
   \hline
   Network Diameter & 14 & 5\\
   \hline
   Page Rank & 0.048138 & 0.041006\\
   \hline
   Graph Density & 0.054 & 0.039\\
   \hline
   Avg clustering coefficient &0.349 &0.275\\
   \hline
   Connected Components &2 &11 \\
   \hline
\end{tabular}}
\end{table} 

\section{Conclusion}
Confinement of an individual during the COVID-19 pandemic, introduced changes to the lifestyle as daily routines were paced down. Food habits and physical activity patterns were investigated for analysing their relationship with changes in body weight and dietary pattern. The study used structured questionnaires to know the awareness of engineering students during lockdown impositions. Fitness-buddy friendship network enabled students to know the pattern and influence of their friends and get motivated towards a healthy lifestyle and fun filled activities. A peer based motivation towards workouts reflects homophily in the classroom.

\bibliographystyle{unsrt}
\bibliography{mybib}

\end{document}